\documentstyle[manuscript,eqsecnum,aps,version2]{revtex}
\input epsf
\draft

\begin{document}
\title
Theory of Abelian Projection
\endtitle
\author{Michael~C.~Ogilvie}
\instit
Department of Physics, Washington University, St. Louis, MO 63130
\endinstit
\medskip
\centerline{\today}

\abstract

Analytic methods for Abelian projection are developed. A
number of results are obtained related to string tension measurements. It is
proven that even without gauge fixing,
abelian projection yields string tensions of the underlying non-Abelian
theory. Strong arguments are given for similar results in the case where
gauge fixing is employed. The methods used emphasize that the projected
theory is derived from the underlying non-Abelian theory rather than \textit{%
vice versa}. In general, the choice of subgroup used for projection is not
very important, and need not be Abelian. While gauge fixing is shown to be
in principle unnecessary for the success of Abelian projection, it is
computationally advantageous for the same reasons that improved operators,
\textit{e.g.}, the use of fat links, are advantageous in Wilson loop
measurements. Two other issues, Casimir scaling and the conflict between
projection and critical universality, are also discussed.

\endabstract


\section{Introduction}

In this paper we study analytically Abelian projection, attempting to
resolve fundamental issues of its utility and interpretation. Abelian
projection is a method for investigating the properties of gauge theories,
particularly those properties associated with confinement. A gauge is chosen
which reduces the gauge symmetry of a non-Abelian group $G$ to its maximal
Abelian subgroup $H$.\cite{'t Hooft} In such a gauge, it
is possible to identify Abelian gauge fields and magnetic monopoles. These
monopoles are presumed to play an essential role in confinement.
\cite{Mandelstam} As a lattice gauge theory technique, Abelian
projection is an algorithm, described below, for extracting an ensemble of
Abelian gauge field configuration from an ensemble of non-Abelian
configurations.

\vspace{1pt}Abelian dominance is a central concept in work on Abelian
projection, positing that the essential non-perturbative physics on the
non-Abelian gauge field is carried in its Abelian projection. A strong form
of Abelian dominance states for every observable $O$ in the non-Abelian
theory, there is a corresponding observable in the Abelian theory $O^{\prime
}$ such that $\left\langle O\right\rangle =\left\langle O^{\prime
}\right\rangle $, where the averages are taken over corresponding ensemble
of non-Abelian and Abelian gauge fields, respectively. Such equalities can
only be approximate at best. However, Abelian projection has had notable
success in relating the string tensions and monopole densities in the
projected theories to related quantities in the underlying non-Abelian
theories.\cite{Suzuki}

\emph{\vspace{1pt}}There are several widely recognized issues having to do
with Abelian projection. \cite{Greensite} The first is
the problem of Casimir scaling. The simplest illustration comes from $SU(2)$%
. Using Abelian dominance, the gauge field in its $3\times 3$ adjoint
representation $(j=1)$ can be decomposed into Abelian gauge fields of charge 
$\pm 1$, with a field identically zero representing the charge $0$ sector.
The corresponding adjoint representation Wilson loop receives a constant
contribution from the charge $0$ sector. Such behavior is in contradiction
to the known area-law behavior of the adjoint Wilson loop at intermediate
distances. This is based on a seductive misapprehension: that the
non-Abelian gauge fields are obtained from the Abelian fields by
''dressing'' them with small non-Abelian fluctuations. In fact, it is the
Abelian fields which are derived from the non-Abelian fields in much the
same way that the renormalization group derives a coarse-grained field from
a fine-grained field. The second problem is the possible triviality of
projection, that it is guaranteed to work for reasons having nothing to do
with Abelian dominance. This work provides support for that position. A
third issue is the correct, or best subgroup to use for projection. As will
be shown below, the success of Abelian projection is largely independent of
the subgroup used. For example, in the case of $SU(3)$, one could use as a
subgroup for projection $U(1)\times U(1)$, $Z(3)$, or even $SU(2)\times U(1)$%
. The next section briefly recapitulates the algorithm for Abelian
projection. Section 3 analyzes projection without gauge fixing, and section
4 projection with gauge fixing. Section 5 discusses Casimir scaling.
Section 6 deals with a conflict between projection and critical
universality which occurs when finite temperature gauge theories are
projected to subgroups other than the center of the gauge group.
A final
section discusses the implications of the results obtained, particularly in
relation to these three issues.

\section{Abelian Projection in Practice}

\vspace{1pt}The standard approach to Abelian projection is a three step
process. The gauge fields are associated with links of the lattice, and
take on values in a compact Lie group $G$. An ensemble of lattice gauge
field configurations is generated using standard Monte Carlo methods. This
ensemble of G-field configurations is generated by a functional integral 
\begin{eqnarray}
Z_{g}=\int \left[ dg\right] e^{S_{g}\left[ g\right] } 
\end{eqnarray}
where $\left[ dg\right] $ will consistently be used to denote the integral
over all fields of a given type; $\left( dg\right) $ will be used for
integrals over individual fields and includes Haar measure. $S_{g}$ is a
gauge-invariant action for the gauge fields, \textit{e.g.}, the Wilson
action for $SU(N)$ gauge fields: 
\begin{eqnarray}
S_{g}=\frac{\beta }{2N}\sum_{plaq}\,Tr\,\left( g_{plaq}+g_{plaq}^{+}\right) 
\end{eqnarray}
where $g_{plaq}$ is a plaquette variable composed from link variables, and
the sum is over all plaquettes of the lattice. The expectation value of any
observable $O$ is given formally by 
\begin{eqnarray}
\left\langle O\right\rangle =\frac{1}{Z_{g}}\int \left[ dg\right]
\,e^{S_{g}\left[ g\right] }\,O 
\end{eqnarray}
but in simulations is evaluated by an average over an ensemble of field
configurations: 
\begin{eqnarray}
\left\langle O\right\rangle =\frac{1}{n}\sum_{i=1}^{n}\,O_{i} 
\end{eqnarray}
Each field configuration in the $G$-ensemble is placed in a particular
gauge. This gauge is chosen to preserve gauge invariance for some subgroup $%
H $ of $G$. For example, when $SU(2)$ is projected to $U(1)$, the gauge
often used is defined by maximizing the quantity 
\begin{eqnarray}
\sum_{l}Tr\,\left[ g_{l}\,\sigma _{3}g_{l}^{+}\sigma _{3}\right] 
\end{eqnarray}
for each configuration over the class of all gauge transformations. The sum
is over all the links of the lattice. This global maximization is often
implemented as a local iterative maximization. While this subgroup has
generally been chosen to be Abelian, I will show later that this is not
essential. From this ensemble of field configurations, another ensemble of
gauge fields is generated from the gauge-fixed ensemble, with the fields
taking on values in the subgroup $H$. This is obtained by maximizing 
\begin{eqnarray}
\sum_{l}Tr\left( g_{l}^{+}h_{l}+h_{l}^{+}g_{l}\right) 
\end{eqnarray}
where $h_{l}\in H$. For example, when $G=SU(2)$ and $H=U(1)$, an element $%
h_{l}$ can be represented as 
\begin{eqnarray}
h_{l}=\left( 
\begin{array}{ll}
e^{i{1 \over 2}\theta _{l}} & 0 \\ 
0 & e^{-i{1 \over 2}\theta _{l}}
\end{array}
\right) 
\end{eqnarray}
using the natural embedding of $U(1)$ in $SU(2)$. Note that the irreducible
representations of $G$ are generally reducible representations of $H$. No
distinction will be made here between an element of $H\,$and its
representation in $G$ for notational convenience, although elsewhere it will
be. The projection procedure can be carried out very efficiently for each
link. It is true, but perhaps not obvious, that the derived ensemble is, in
the limit of large numbers of configurations $n$, invariant under gauge
transformations in $H$.

For analytical purposes, it is necessary to generalize this procedure, so
that a given single configuration of $G$-fields will be associated with an
ensemble of configurations of $H$-fields. We will generate this ensemble
using 
\begin{eqnarray}
S_{proj}\left[ g,h\right] =\sum_{l}\left[ \frac{p}{2N}Tr\left(
g_{l}^{+}h_{l}+h_{l}^{+}g_{l}\right) \right] 
\end{eqnarray}
as a weight function to select an ensemble of $h$ fields. The normal
procedure is formally regained in the limit $p\rightarrow \infty $.
Computationally, this would be implemented as a Monte Carlo simulation
inside a Monte Carlo simulation. Note that the $h$ fields should be thought
of as quenched variables, since they do not effect the $g$-ensemble. We will
treat the gauge fixing process similarly later.

\section{Projection without Gauge Fixing}

We begin with a discussion of projection without gauge fixing, as this is
the simplest and most analytically tractable case. As discussed above, it is
necessary for analytical purposes to generalize the projection procedure, so
that a given configuration of $G$-fields will be associated with an ensemble
of configurations of $H$-fields. This can be done by extending the
definition of the expectation value of an operator $O$ as 
\begin{eqnarray}
\left\langle O\right\rangle =\frac{1}{Z_{g}}\int \left[ dg\right]
\,e^{S_{g}\left[ g\right] }\frac{1}{Z_{proj}\lbrack g\rbrack }\int \left[
dh\right] e^{S_{proj}\left[ g,h\right] }\,O 
\end{eqnarray}
where the presence of $Z_{proj}\lbrack g\rbrack $, defined as 
\begin{eqnarray}
Z_{proj}\lbrack g\rbrack =\int \left[ dh\right] \,e^{S_{proj}\left[
g,h\right] } 
\end{eqnarray}
is crucial. If $O$ depends only on the $g$ fields, this definition reduces
to the previous case. $Z_{proj}\lbrack g\rbrack $ ensures that the $h$
fields behave as quenched variables, and have no effect on the distribution
of $g$ fields. It may be helpful to compare the $h$ fields to the role of
quark fields in the quenched approximation of lattice QCD. The role of $%
Z_{proj}\lbrack g\rbrack $ is analogous to the fermion determinant. It is
easy to check that if the observable $O$ is invariant under gauge transforms
in $H$, so is its expectation value $\left\langle O\right\rangle $. Note
that $Z_{proj}\lbrack g\rbrack $ is invariant under gauge transformations in 
$H$ acting on $g$.

A crucial tool in our analysis will be the character expansion.\cite{Creutz}
\cite{Itzykson}
Each
irreducible representation of $G$ or $H$ will have a label $\alpha $, so
that $D_{ij}^{\alpha }(g)$ is the $i,\,j$ entry of a matrix representative
of $g$ in the irreducible representation $\alpha $. Irreducible
representations of $H$ will be denoted by $\widetilde{D}_{ij}^{\alpha }(h)$.
The fundamental orthogonality relation for matrix elements is 
\begin{eqnarray}
\int_{G}(dg)\;D_{ij}^{\alpha }(g)D_{kl}^{\beta }(g^{+})=\frac{1}{d_{\alpha }}%
\delta _{\alpha \beta }\,\delta _{il}\,\delta _{jk}
\end{eqnarray}
where the integral is over Haar measure, conventionally normalized to $1$,
and $d_{\alpha }$ is the dimensionality of the irreducible representation $%
\alpha $. The group character $\chi ^{\alpha }(g)$ is the trace of a group
element in a particular representation: $\chi ^{\alpha }(g)=D_{ii}^{\alpha
}(g),$ where the summation convention is employed. It follows immediately
that 
\begin{eqnarray}
\int_{G}(dg)\;\chi ^{\alpha }(g)\chi ^{\beta }(g^{+})=\delta _{\alpha \beta }
\end{eqnarray}

The weight function for projecting each link can be expanded in the
characters of the group $G$, 
\begin{eqnarray}
\exp \left[ \frac{p}{2N}Tr\left( g^{+}h+h^{+}g\right) \right] =\sum_{\alpha
}d_{\alpha }c_{\alpha }\left( p\right) \chi _{\alpha }\left( h^{+}g\right) 
\end{eqnarray}
in a generalization of Fourier decomposition. The coefficients of the
expansion are given by 
\begin{eqnarray}
c_{\alpha }(p)=\frac{1}{d_{\alpha }}\int_{G}\,(dg)\,\chi _{\alpha }\left(
g^{+}\right) \exp \left[ \frac{p}{2N}Tr\left( g+g^{+}\right) \right] 
\end{eqnarray}
and are known for several common groups. For example, in the case of $SU(2)$%
, the characters are labeled by a non-negative integer or half-integer $j$,
the dimensionality $d_{j}=2j+1$, and the coefficients $c_{j}$ are given by $%
c_{j}(p)=2\,I_{2j+1}(p)/p$.

It is easy to see that $Z_{proj}\lbrack g\rbrack $ is given by 
\begin{eqnarray}
Z_{proj}\lbrack g\rbrack =\prod_{l}\,\widetilde{c}_{0}(p,g_{l}) 
\end{eqnarray}
where $\widetilde{c}_{0}$ is given by 
\begin{eqnarray}
\widetilde{c}_{0}(p,g_{l})=\int_{H}(dh_{l})\,\exp \left[ \frac{p}{2N}%
Tr\left( g_{l}h_{l}+h_{l}^{+}g^{+}\right) \right] 
\end{eqnarray}
If $H$ is abelian, the weight function can also be expanded in characters of
the subgroup $H$:

\begin{eqnarray}
\exp \left[ \frac{p}{2N}Tr\left( g^{+}h+h^{+}g\right) \right] =\sum_{\alpha }%
\widetilde{d}_{\alpha }\widetilde{c}_{\alpha }(p,g)\widetilde{\chi }_{\alpha
}\left( h\right) 
\end{eqnarray}
where the coefficients $\widetilde{c}_{\alpha }(p,g)$ depend on $g$. For
example, in the case $G=SU(2)$ and $H=Z(2)$, we have 
\begin{eqnarray}
\exp \left[ \frac{p}{2}Tr\left( zg\right) \right] =
1/2~ \cosh ( p\,Tr\,g )
+1/2~ z \sinh ( p\,Tr\,g )
\end{eqnarray}
making clear that the projection weight is not invariant under gauge
transformation in $G$.

We now examine the computation of the expectation value of a Wilson loop $W$
which has no self-intersections; in particular consider 
\begin{eqnarray}
\widetilde{\chi }^{\beta }(h_{1}..h_{n})=\widetilde{D}_{j_{1}\,j_{2}}^{\beta
}(h_{1})\widetilde{D}_{j_{2}\,j_{3}}^{\beta }(h_{2})..\widetilde{D}%
_{j_{n\,}\,j_{1}}^{\beta }(h_{n})
\end{eqnarray}
Each term in the product will be paired with terms from the character
expansion of the projection weight for that link. A typical term has the
form 
\begin{eqnarray}
\,\widetilde{D}_{j_{m}\,j_{m+1}}^{\beta }(h_{m})\sum_{\alpha }d_{\alpha
}c_{\alpha }\chi _{\alpha }\left( h_{m}^{+}g_{m}\right) =\,\widetilde{D}%
_{j_{m}\,j_{m+1}}^{\beta }(h_{m})\sum_{\alpha }d_{\alpha }c_{\alpha
}D_{k_{m}\,l_{m}}^{\alpha }\left( h_{m}^{+})D_{l_{m}\,k_{m}}^{\alpha
}(g_{m}\right) 
\end{eqnarray}
where $m$ is a particular link and $\beta $ is an index associated with an
irreducible representation of $H$. At this point, we invoke the gauge
invariance of the underlying theory and the non-intersecting character of
the curve $W$. For the moment, we also set $Z_{proj}\lbrack g\rbrack $ equal
to its lowest-order expression in the expansion in characters of $G$: 
\begin{eqnarray}
Z_{proj}\lbrack g\rbrack =\prod_{l}\,c_{0}(p)
\end{eqnarray}
Consider two adjacent links on the curve $g_{m}$ and $g_{m+1}$. We are free
to make a change of variable on all the links associated with their common
site which has the form of a gauge transformation: $g_{m\,}$is replaced by $%
g_{m\,}\phi_{m}$ and $g_{m+1\,}$by $\phi_{m}^{+}g_{m+1\,}$and so forth in such a
way that the action $S_{g}$ is left invariant. We are free to integrate over
the variable $\phi_{m}$\thinspace , giving 
\begin{eqnarray}
\int_{G}(d\phi_{m})D_{l_{m}\,k_{m}}^{\alpha
_{m}}(g_{m}\phi_{m})D_{l_{m+1}\,k_{m+1}}^{\alpha _{m+1}}(\phi_{m}^{+}g_{m+1})=%
\frac{1}{d_{\alpha _{m}}}\delta _{\alpha _{m}\alpha _{m+1}}\delta
_{k_{m}l_{m+1}}D_{l_{m}k_{m+1}}^{\alpha _{m}}(g_{m}g_{m+1})
\label{eqn0}
\end{eqnarray}
Systematic application of this result at all sites along the curve $W$
collapses the sum into the simple result 
\begin{eqnarray}
\left\langle \widetilde{\chi }^{\beta }(h_{1}..h_{n})\right\rangle
=\sum_{\alpha }\left( \frac{c_{\alpha }(p)}{c_{0}(p)}\right)
^{n}\int_{H}(dh)\,\widetilde{\chi }^{\beta }(h)\chi ^{\alpha
}(h^{+})\,\,\left\langle \chi ^{\alpha }(g_{1}..g_{n})\right\rangle 
\end{eqnarray}
The single integral over $h$ occurs because the integral over all the fields 
$h_{1}..h_{n}$, which has the form
\begin{eqnarray}
\int_{H}(dh_{1})..(dh_{n})\,\widetilde{\chi }^{\beta }(h_{1}..h_{n})\chi
^{\alpha }(h_{n}^{+}..h_{1}^{+})
\end{eqnarray}
can be simplified by the change of variable $h=h_{1}..h_{n}$. This integral
returns a non-negative integer which is the number of times the
representation $\beta $ of $H$ is contained in the representation $\alpha $
of $G$. While this result for the Wilson loop in the $\beta \,$%
representation of $H$ treats the numerator exactly, it is only \ the
lowest-order approximation to the denominator. However, it has obvious
physics content: the Wilson loop as measured in the $\beta $ representation
of $H$ is given as a sum of Wilson loops in the irreducible representations
of $G$, each weighted by the number of times $\beta $ appears in $\alpha $
and by a $p$-dependent factor which contribute to the perimeter dependence
and goes to one as $p\rightarrow \infty $. This result will be the starting
point for all subsequent cases considered.

We can turn this approximate result into rigorous upper and lower bounds, by
noting that $\widetilde{c}_{0}(p,g_{m})$ is bounded: 
\begin{eqnarray}
e^{pM_{1}}\leq \widetilde{c}_{0}(p,g_{m})\leq e^{pM_{2}}
\end{eqnarray}
where $M_{1}$ and $M_{2}$ are the lower and upper bounds of the projection
function. Thus we have the upper bound 
\begin{eqnarray}
\left\langle \widetilde{\chi }^{\beta }(h_{1}..h_{n})\right\rangle \leq
\sum_{\alpha }\left( c_{a}(p)e^{-pM_{1}}\right) ^{n}\int_{H}(dh)\,\widetilde{%
\chi }^{\beta }(h)\chi ^{\alpha }(h^{+})\,\,\left\langle \chi ^{\alpha
}(g_{1}..g_{n})\right\rangle 
\end{eqnarray}
with a corresponding lower bound when $M_{1}$ is replaced by $M_{2}$. These
bounds hold for $0<p<\infty $. The lower bound requires two extra
assumptions. First is the Griffiths-type inequality $\left\langle \chi
^{\alpha }(g_{1}..g_{n})\right\rangle \geq 0$, which has not been proven for
non-Abelian gauge theories, but should hold on physical grounds for
rectangular Wilson loops. Second is the assumption that the coefficients $%
c_{a}(p)$ are non-negative; this holds for the projection function
considered here, but might fail for others. Assuming that all the
representations contributing to the sum have area law behavior, we have the
result, independent of $p$, 
\begin{eqnarray}
\tilde{\sigma}_{\beta }=\min_{\alpha }\,\sigma _{a}
\end{eqnarray}
where the minimum is taken over all representations $\alpha $ that have a
non-zero contribution. This is the key result for this section.

Consider, as an example, the case of $SU(2)$ projected to $U(1)$. The string
tension is non-zero for the half-integer representations $j=1/2,\,3/2,\,..$
due to the $Z_{2}$ center symmetry, but not for the integer representations.
Because $Z_{2}\subset U(1)$, the integral 
\begin{eqnarray}
\int_{U(1)}(dh)\,\widetilde{\chi }^{n}(h)\chi ^{j}(h^{+})\, 
\end{eqnarray}
will vanish in many cases. Typical equalities include 
\begin{eqnarray}
\tilde{\sigma}_{1/2}=\min_{j=1/2,\,3/2,\,..}\,\sigma _{j} 
\end{eqnarray}
and 
\begin{eqnarray}
\tilde{\sigma}_{3/2}=\min_{j=3/2,\,5/2,\,..}\,\sigma _{j} 
\end{eqnarray}
but $\tilde{\sigma}_{1}=0$ because of string-breaking in the adjoint
representation: $\sigma _{1}=0$. Note that $U(1)$ charges have been
normalized such that the highest $U(1)$ charge associated with an $SU(2)$
representation $j$ \ has charge given by $m=j$ so that $U(1)$
integrations must go over $4 \pi$.

In the case of $SU(N)$ projected to $Z(N)$, a very direct alternative
proof has been given recently by Ambjorn and Greensite.\cite{Ambjorn}
They observe that the the $Z(N)$ projection of an $SU(N)$ matrix
is a class function, and thus may be directly expanded in
characters. For example, the $Z(2)$ projection of an $SU(2)$ element
is given simply as $sign ( Tr U )$, which is readily expanded in
a character expansion. This yields exact relations between $Z(N)$
$SU(N)$ Wilson loops.

There are obvious similarities between projection and renormalization group
transformations. In particular, the projection function is also used in real
space renormalization group calculations, where it is used to project blocks
of spins back onto the space of site variables. For example, in the case of
an $SO(N)\;$spin model where the spin variable $\sigma $ is an $N$-dimensional
real unit vector, the weight function for a typical real space
renormalization group transformation would be
\begin{eqnarray}
\exp \left[ \frac{p}{N}\sigma ^{^{\prime }}\cdot \left( \sum_{block}\sigma
\right) \right] 
\end{eqnarray}
where the sum of all the spins $\sigma $ in a block are projected back to a
spin variable $\sigma ^{^{\prime }}$. In view of this connection, it is
interesting to take $H=G$ so that the subgroup is in fact the group. In that
case, we have 
\begin{eqnarray}
\widetilde{c}_{0}(g)=\int_{G}(dh)\,\exp \left[ \frac{p}{2N}Tr\left(
g^{+}h+h^{+}g\right) \right] =\int_{G}(dh)\,\exp \left[ \frac{p}{2N}Tr\left(
h+h^{+}\right) \right] =c_{0}(p)
\end{eqnarray}
giving the exact result 
\begin{eqnarray}
\left\langle \widetilde{\chi }^{\alpha }(h_{1}..h_{n})\right\rangle =\left( 
\frac{c_{\alpha }(p)}{c_{0}(p)}\right) ^{n}\,\,\left\langle \chi ^{\alpha
}(g_{1}..g_{n})\right\rangle 
\end{eqnarray}
showing that the only effect of projection in this case is a finite
renormalization of the perimeter contribution. This corresponds to field
renormalization for renormalization group transformations. Note that the
perimeter renormalization is always $\leq 1$, attaining $1$ in the limit $%
p\rightarrow \infty $. This is easily understood as a consequence of the
smearing properties of the transformation.

\section{Projection with Gauge Fixing}

Projection combined with gauge fixing is much more difficult to analyze. We
assume we are given some gauge fixing function $S_{gf}\lbrack g\rbrack $ to
maximize which, while not invariant under local gauge transformations in $G$%
, is invariant under local gauge transformations in $H$. In the case of $%
SU(2)$ projected to $U(1)$, we take 
\begin{eqnarray}
S_{gf}=\lambda \sum_{l}Tr\,\left[ g_{l}\,\sigma _{3}g_{l}^{+}\sigma
_{3}\right] 
\end{eqnarray}
where the parameter $\lambda $ has been introduced in the same way $p$ was
introduced for projection. It is very convenient to introduce an auxiliary
gauge-fixing field $\phi (x)$, which takes values in $G$. This fields are
applied to an unfixed field configuration as $g_{\mu }(x)\rightarrow 
\widetilde{g}_{\mu }(x)=\phi (x)g_{\mu }(x)\phi ^{+}(x+\widehat{\mu })$ so
that $\widetilde{g}_{\mu }(x)$ is used wherever the gauge-fixed field is
required. The gauge fixing function depends on $\widetilde{g}$, which is to
say both $g$ and $\phi $. The expectation value of an observable $O$ is now
given by 
\begin{eqnarray}
\left\langle O\right\rangle =\frac{1}{Z_{g}}\int \left[ dg\right]
\,e^{S_{g}\left[ g\right] }\frac{1}{Z_{gf}\lbrack g\rbrack }\int \left[
d\phi \right] e^{S_{gf}\left[ \widetilde{g}\right] }\frac{1}{Z_{proj}\lbrack 
\widetilde{g}\rbrack }\int \left[ dh\right] e^{S_{proj}\left[ \widetilde{g}%
,h\right] }\,O
\end{eqnarray}
where $Z_{gf}\lbrack g\rbrack $, defined as 
\begin{eqnarray}
Z_{gf}\lbrack g \rbrack =\int \left[ d\phi \right]
\,e^{S_{gf}\left[ \widetilde{g}\right] }
\end{eqnarray}
is needed for the same reason that $Z_{proj}$ was before.\cite{Fachin}
In the limit $%
\lambda \rightarrow \infty $, the procedural implementation of this formula
is equivalent to the commonly used
algorithm for lattice gauge fixing.\cite{Mandula}
As before, it is easy to check that invariance
under gauge transformations in $H$ holds.

Formally, the field $\phi $ is just a quenched, adjoint representation
scalar field. It has two independent local symmetry groups: $H_{L}\otimes
G_{R}$. The generating function $Z_{gf}\lbrack g\rbrack $ is a lattice
analog of the Fadeev-Popov determinant (actually the inverse of the
determinant). However, there are some important differences. Note
immediately that $Z_{gf}\lbrack g\rbrack $ depends on the gauge-fixing
parameter $\lambda $. More fundamentally, with the continuum Fadeev-Popov
determinant, there is the vexing question of Gribov copies: what should be
done about field configurations on the same gauge orbit satisfying the same
gauge condition? The lattice formalism avoids this question. By
construction, gauge-invariant observables are evaluated by integrating over
all configurations. Gauge-variant quantities receive weighted contributions
from Gribov copies. Thus the connection between lattice gauge fixing and
gauge fixing in the continuum is not simple.

We have not been able to derive general results similar to those obtained
without gauge fixing. However, there are strong reasons for believing that
similar results hold in the gauge-fixed case as well.

First, rigorous bounds will be proven for one-dimensional gauge-fixing,
where the gauge-fixing condition only depends on the links in one direction,
which we take to be the timelike direction. An example of such a gauge, in
an obvious notation, is 
\begin{eqnarray}
S_{gf}=\lambda \sum_{\vec{x},t} Re~Tr\,\left[ g_{0}
( \vec{x},t)\right] 
\end{eqnarray}
which is maximized when all the links associated with a given spatial value $%
\vec{x}$ are equal. This particular gauge does not have a
gauge-invariant subgroup $H$ associated with it, but other gauges which do
can also be used, for example 
\begin{eqnarray}
S_{gf}=\lambda \sum_{\vec{x},t}Tr\,\left[ g_{0}(\vec{x}%
,t)\,Mg_{0}^{+}(\vec{x},t)M\right] 
\end{eqnarray}
where M is a Hermitian matrix. The matrix $M$ can be taken to be traceless,
since the trace would only contribute a constant. The set of group elements
that commute with $M$ determines the subgroup $H$. In the case of
one-dimensional gauge fixing, it is natural to use a lattice at finite
temperature, taken to be $T$, and to use Polyakov loops instead of Wilson
loops as observables. Note that $T\,$can be taken arbitrarily large, so the
restriction to finite temperature is not significant. Consider a correlation
function of the form 
\begin{eqnarray}
\left\langle \widetilde{\chi }^{\beta }(P(\vec{x_{1}}%
))\right\rangle =\left\langle \widetilde{\chi }^{\beta }(h_{0}(%
\vec{x}_{1},1)..h_{0}(\vec{x}_{1},T))\right\rangle 
\end{eqnarray}
which is the expectation value of a Polyakov loop. The one dimensional
character of the gauge-fixing greatly simplifies the integrals over $h$ and $%
\phi $. Let $M_{3}$ and $M_{4}\,$be lower and upper bounds for each link's
contribution to the gauge-fixing functional $S_{gf}$, with $M_{1}$ and $M_{2}
$ remaining as similar bounds for the projection function. Then it is easy
to see that 
\begin{eqnarray}
\left| \left\langle \widetilde{\chi }^{\beta }(P(\vec{x_{1}}%
)\right\rangle \right|  &\leq &\sum_{\alpha }c_{\alpha
}(p)^{T}e^{-T\,\,pM_{1}+T\,\lambda \left( M_{4}-M_{3}\right) }\int_{H}(dh)\,%
\widetilde{\chi }^{\beta }(h)\chi ^{\alpha }(h^{+})\cdot  
\,\,\left| \left\langle \chi ^{\alpha }(P(\vec{x_{1}}%
)\right\rangle \right| 
\end{eqnarray}
which establishes that the representation $\beta $ of $H$ is confined if all
the representations $\alpha $ in $G$ which contain it are confined. If $%
\beta $ transforms non-trivially under the center of $G$, then $\beta $ is
confined when the center symmetry of $G$ is unbroken,
a result which can also be
obtained on the basis of center symmetry alone. There is no lower bound in
this case, because the integrand in the numerator is non-positive and
difficult to approximate. For Polyakov loop two point functions, defined by 
\begin{eqnarray}
\left\langle \widetilde{\chi }^{\beta }(P(\vec{x_{1}}))\widetilde{%
\chi }^{\beta }(P^{+}(\vec{x}_{2}))\right\rangle =\left\langle 
\widetilde{\chi }^{\beta }(h_{0}(\vec{x}_{1},1)..h_{0}(%
\vec{x}_{1},T))\widetilde{\chi }^{\beta }(h_{0}(\vec{x}%
_{2},1)..h_{0}(\vec{x}_{2},T))\right\rangle 
\end{eqnarray}
a similar upper bound can be obtained: 
\begin{eqnarray*}
\left| \left\langle \widetilde{\chi }^{\beta }(P(\vec{x_{1}}))%
\widetilde{\chi }^{\beta }(P^{+}(\vec{x}_{2}))\right\rangle
\right|  &\leq &\sum_{\alpha ,\gamma }c_{\alpha }(p)^{T}c_{\gamma
}(p)^{T}e^{-2T\,\,pM_{1}+2T\,\lambda \left( M_{4}-M_{3}\right)
}\int_{H}(dh)\,\widetilde{\chi }^{\beta }(h)\chi ^{\alpha }(h^{+})\cdot  \\
&&\,\,\int_{H}(dh)\,\widetilde{\chi }^{\beta }(h^{+})\chi ^{\gamma
}(h)\left| \left\langle \chi ^{\alpha }(P(\vec{x_{1}}))\chi
^{\gamma }(P^{+}(\vec{x}_{2}))\right\rangle \right| .
\end{eqnarray*}
If the original two point function has confining behavior, \textit{i.e.}, 
\begin{eqnarray}
\left\langle \widetilde{\chi }^{\beta }(P(\vec{x_{1}}))\widetilde{%
\chi }^{\beta }(P^{+}(\vec{x}_{2}))\right\rangle \sim \exp \left[
-\widetilde{\sigma }_{\beta }T\left| \vec{x}_{1}-\vec{x%
}_{2}\right| \right] 
\end{eqnarray}
then, using the same arguments as in the previous section, 
\begin{eqnarray}
\tilde{\sigma}_{\beta }\geq \min_{\alpha ,\gamma }\,\sigma _{a\gamma }
\end{eqnarray}
where, as before, the minimum is taken over all representations $\alpha $
that have a non-zero contribution and $\sigma _{a\gamma }$ denotes the
string tension measured by the mixed correlation function. Such bounds are
useful in the case of one-dimensional gauge fixing because bounds on the
gauge fixing term grow only exponentially with $T$, renormalizing perimeter
terms. Unfortunately, they do not seem to extend to higher dimension.

In fact, it is very likely that the equality proved in the non-gauge-fixed
case also holds for the case where gauge-fixing is used. As will now be
shown, strong-coupling expansions in $\lambda $ indicate very little
difference with the cases already considered. The strong-coupling expansion
is convergent for sufficiently small $\lambda $. As before, \vspace{1pt}we
consider the computation of the expectation value of a Wilson loop $W$ which
has no self-intersections: 
\begin{eqnarray}
\widetilde{\chi }^{\beta }(h_{1}..h_{n})=\widetilde{D}_{j_{1}\,j_{2}}^{\beta
}(h_{1})\widetilde{D}_{j_{2}\,j_{3}}^{\beta }(h_{2})..\widetilde{D}%
_{j_{n\,}\,j_{1}}^{\beta }(h_{n}) 
\end{eqnarray}
Each term in the product will be paired with terms from the character
expansion of the projection weight for that link. A typical term has the
form 
\begin{eqnarray}
\,\widetilde{D}_{j_{m}\,j_{m+1}}^{\beta }(h_{m})\sum_{\alpha }d_{\alpha
}c_{\alpha }\chi _{\alpha }\left( h_{m}^{+}\phi _{m}g_{m}\phi
_{m+1}^{+}\right) 
\end{eqnarray}
where $m$ is a particular link and $\beta $ is an index associated with an
irreducible representation of $H$. At order $\lambda ^{0}$, the same
argument given for the ungauge-fixed case applies.
Integration over the $\phi $
fields yields the same result as before: 
\begin{eqnarray}
\left\langle \widetilde{\chi }^{\beta }(h_{1}..h_{n})\right\rangle
=\sum_{\alpha }\left( \frac{c_{\alpha }(p)}{c_{0}(p)}\right)
^{n}\int_{H}(dh)\,\widetilde{\chi }^{\beta }(h)\chi ^{\alpha
}(h^{+})\,\,\left\langle \chi ^{\alpha }(g_{1}..g_{n})\right\rangle 
\end{eqnarray}
where terms of higher order in $p$ and $\lambda $ have been neglected.

Higher order terms in $\lambda $ lead to corrections of this basic result,
which should not alter the basic behavior. For simplicity, consider a
typical contribution in the fundamental representation of $SU\left( N\right) 
$ with $N>2$. The latter restriction eliminates some graphs which are
special to $SU(2)$. Let the gauge fixing function have the form 
\begin{eqnarray}
S_{gf}=\lambda \sum_{l}Tr\,\left[ g_{l}\,Mg_{l}^{+}M\right] 
\end{eqnarray}
where, as before, $M$ is a traceless, Hermitian matrix that commutes with
every element of the subgroup $H$ and the sum is taken over all links.

Consider a straight segment which contributes to $\left\langle \chi \left(
h_{1}..h_{n}\right) \right\rangle $: 
\begin{eqnarray}
Tr\,h_{1}^{+}\phi _{a}g_{1}\phi _{b}^{+}\,Tr\,h_{2}^{+}\phi _{b}g_{2}\phi
_{c}^{+}\,Tr\,h_{3}^{+}\phi _{c}g_{3}\phi _{d}^{+}
\end{eqnarray}
Integration over $\phi _{b}$ and $\phi _{c}$ gives 
\begin{eqnarray}
\left( \frac{1}{N}\right) ^{2}\,Tr\,\phi _{a}g_{1}g_{2}g_{3}\phi
_{d}h_{3}^{+}h_{2}^{+}h_{1}^{+}
\end{eqnarray}
which is a piece of the $O\left( \lambda ^{0}\right) $ result. We now
consider a $O(\lambda ^{3})$ correction of the form: 
\begin{eqnarray*}
&&Tr\,h_{1}^{+}\phi _{a}g_{1}\phi _{b}^{+}\,Tr\,h_{2}^{+}\phi _{b}g_{2}\phi
_{c}^{+}\,Tr\,h_{3}^{+}\phi _{c}g_{3}\phi _{d}^{+}\,\cdot  \\
&&Tr\,\phi _{b}g_{A}\phi _{e}^{+}M\phi _{e}g_{A}^{+}\phi _{b}^{+}M\;Tr\,\phi
_{e}g_{B}\phi _{f}^{+}M\phi _{f}g_{B}^{+}\phi _{e}^{+}M\;Tr\,\phi
_{f}g_{C}\phi _{c}^{+}M\phi _{c}g_{C}^{+}\phi _{f}^{+}M
\end{eqnarray*}
where integrations must now be performed over $G$ for the fields $\phi _{b}$%
, $\phi _{c}$, $\phi _{e}$, and $\phi _{f}$. Figure 1 shows the labeling of
the sites and links. It is quite worthwhile to use graphical techniques for
evaluating these integrals. Using the techniques developed by Creutz
\cite{Creutz2} and a large number of colored pens, the result is 
\begin{eqnarray*}
&&\frac{\lambda ^{3}\left( Tr\,M^{2}\right) ^{2}}{\left( N^{2}-1\right) ^{4}}%
Tr\,\left[ \phi _{a}g_{1}g_{A}g_{B}g_{C}g_{3}\phi
_{d}h_{3}^{+}h_{2}^{+}h_{1}^{+}M^{2}\right] \,Tr\left[
g_{2}g_{C}^{+}g_{B}^{+}g_{A}^{+}\right] + \\
&&-\frac{\lambda ^{3}\left( Tr\,M^{2}\right) ^{2}}{N\left( N^{2}-1\right)
^{4}}\,Tr\,\phi _{a}g_{1}g_{2}g_{3}\phi _{d}h_{3}^{+}h_{2}^{+}h_{1}^{+}M^{2}
\end{eqnarray*}
Up to multiplicative coefficients and commutation of $M\,$with $h$, this
form is uniquely required by the $H_{L}\otimes G_{R}$ gauge invariance. From
this, we can see that the expectation value $\left\langle \chi \left(
h_{1}..h_{n}\right) \right\rangle $ will be given by 
\begin{eqnarray*}
\left\langle \chi \left( h_{1}..h_{n}\right) \right\rangle  &\approx
&p^{n}K_{1}\left( p,\lambda ,M,N\right) \left\langle \chi \left(
g_{1}..g_{n}\right) \right\rangle  \\
&&+p^{n}\lambda ^{3}K_{2}\left( p,\lambda ,M,N\right)
\sum_{all~insertions} \left\langle \chi \left(
g_{1}g_{A}g_{B}g_{C}..g_{n}\right) \chi \left(
g_{2}g_{C}^{+}g_{B}^{+}g_{A}^{+}\right) \right\rangle 
\end{eqnarray*}
where the summation sign indicates that the same sort of insertion performed
at the one link is to be repeated through the entire loop for all directions
orthogonal to the loop. The constants $K_{1}$ and $K_{2}$ are power series
in $p$ and $\lambda $, beginning at order $1$. This formula is not exactly
correct, because there are $O\left( \lambda ^{2}\right) $ terms associated
with corners. However, it does capture the lowest-order corrections to the
area- and perimeter-dependence. These corrections are shown graphically in
Figure 2. From this expression, we can see that the string tension should
still satisfy 
\begin{eqnarray}
\tilde{\sigma}_{\beta }=\min_{\alpha }\,\sigma _{a}
\end{eqnarray}
as in the case of no gauge fixing. The parameters $p$ and $\lambda $ control
the perimeter dependence of the Wilson loop expectation value, but not the
area dependence.

It is interesting to compare gauge fixing with the use of fat links, which
leads to a similar result. Let the projection function be 
\begin{eqnarray}
S_{proj}\left[ g,h\right] =\sum_{l}\left[ \frac{p}{2N}Tr\left( \overline{g}%
_{l}^{+}h_{l}+h_{l}^{+}\overline{g}_{l}\right) \right] 
\end{eqnarray}
where $\overline{g_{l}}$ is a fat link defined by 
\begin{eqnarray}
\overline{g}_{\mu }(x)=g_{\mu }(x)+\gamma \sum_{\nu \neq \mu }g_{\nu
}(x)g_{\mu }(x+\widehat{\nu })g_{-\mu }(x+\widehat{v}+\widehat{\mu }) 
\end{eqnarray}
where $\gamma $ is an arbitrary real number, generally taken to be positive.
It is now necessary to calculate perturbatively in $\gamma $ the corrections
to the lowest order result, which are shown in Figure 3.

In both cases we have considered, the corrections to the lowest order result
have been slightly different, but all should give the same result for the
string tension. The perimeter dependence will be different for each case,
and depends on the projection parameter $p$, as well as the gauge fixing
parameter $\lambda $ or any other parameters involved. In practice, some
schemes will be numerically advantageous. It is likely that extraction of
the string tension without gauge fixing would be highly inefficient, unless
fat links or some equivalent were used. Fundamentally, it appears that gauge
fixing is providing the same kind of advantage that fat links do, by
providing an improved operator for measuring the string tension, constructed
implicitly by the gauge fixing procedure.

\vspace{1pt}

\section{Casimir Scaling}

The intuitive picture of Abelian dominance, if not Abelian projection, is
that the dominant contribution to the partition function comes from Abelian
field configurations dressed by non-Abelian fluctuations. In this kind of
picture, it makes sense to conjecture that for $SU(2)$ projected to $U(1)$%
\begin{eqnarray}
\left\langle \chi _{j=1}\right\rangle \sim \left\langle \widetilde{\chi }%
_{m=1}\right\rangle +\left\langle \widetilde{\chi }_{m=-1}\right\rangle
+\left\langle \widetilde{\chi }_{m=0}\right\rangle 
\end{eqnarray}
where needed multiplicative coefficients are suppressed for notational
simplicity. However, as has been shown above, this is not the sort of
relation which naturally emerges. The natural relations give $U(1)$ Wilson
loops in terms of $SU(2)$ Wilson loops: 
\begin{eqnarray*}
\left\langle \widetilde{\chi }_{m=0}\right\rangle  &=&\left\langle \chi
_{j=0}\right\rangle  \\
\left\langle \widetilde{\chi }_{m=1}\right\rangle  &\sim &\left\langle \chi
_{j=1}\right\rangle +\left\langle \chi _{j=2}\right\rangle +.. \\
\left\langle \widetilde{\chi }_{m=2}\right\rangle  &\sim &\left\langle \chi
_{j=2}\right\rangle +\left\langle \chi _{j=3}\right\rangle +..
\end{eqnarray*}
where coefficients have again been suppressed for notational simplicity and
only the integer charges, which transform trivially under $Z(2)$ are shown.
Also suppressed are the notational complexities arising from the appearance
of more complicated loops. Notice that the $n=0$ relation is exact. I have
assumed that there is no $j=1$ contribution to $n=4$, for example, but that
is inessential. More importantly, there should be no $j=0$ contribution for $%
n>0$, since that would give a constant contribution, rather than the
expected area or perimeter behavior. If that is the case, we can imagine
inverting these relations, obtaining, for example 
\begin{eqnarray}
\left\langle \chi _{j=1}\right\rangle \sim \left\langle \widetilde{\chi }%
_{m=1}\right\rangle +\left\langle \widetilde{\chi }_{m=2}\right\rangle +..
\end{eqnarray}
where no $m=0$ contribution appears. This would resolve the issue of Casimir
scaling discussed in the introduction, by establishing that there is no $%
n=0\,$contribution to $j=1$. As an example, it is possible to invert 
equation \ref{eqn0}, yielding. 
\begin{eqnarray}
\left\langle \chi _{j}\right\rangle =\frac{1}{2}\left( \frac{c_{0}(p)}{%
c_{j}(p)}\right) ^{n}\left[ \left\langle \widetilde{\chi }%
_{m=j}\right\rangle -\left\langle \widetilde{\chi }_{m=j+1}\right\rangle
\right] 
\end{eqnarray}
valid for $j>0$. While this equation should not be taken too seriously, it
does show that the problem of Casimir scaling may be resolved in a simple
way.

\section{\protect\vspace{1pt}Critical Behavior and Universality}

There is a puzzling aspect of Abelian projection associated with
universality, which is not immediately apparent when considering
only $SU(N)$ gauge theories, but appears immediately when
considering the analogous procedure in spin models.
In a generic spin model, the spin $\sigma$ takes on values
in some space $M$. A group $G$ acts on $\sigma$ in such a way that
the Hamiltonian of the spin system is invariant. Imagine
projecting $\sigma$ to a new spin variable $\mu$ which lies
in a subspace $N$ of $M$; the new variable $\mu$ will have a
symmetry group $H$ which is a subgroup of $G$. Obviously, any
critical behavior of the original system will be reflected in
the behavior of the projected variables.

For the sake of concreteness, take $\sigma \in S^{N-1}$,
the unit sphere in $N$ dimensions, and the
symmetry group to be $O(N)$. Take $\mu \in Z(2)$.
The projection function can be taken to be
\begin{eqnarray}
\sum_{s} p \mu_s (e \cdot \sigma )
\end{eqnarray}
where $e$ is a fixed element of $S^{N-1}$ and $p$ is an adjustable
parameter as before.
When considering spontaneous
symmetry breaking where the direction of the field is determined by
an infinitesimal symmetry breaking field or by boundary conditions,
it is natural to choose $e$ in the appropriate direction.
It is easy to derive relations between the correlation functions,
{\it e.g.},
\begin{eqnarray}
\left\langle \mu_i \right\rangle =
\left\langle tanh ( p e \cdot \sigma_i ) \right\rangle
\end{eqnarray}
and
\begin{eqnarray}
\left\langle \mu_i \mu_j \right\rangle =
\left\langle tanh ( p e \cdot \sigma_i )
tanh ( p e \cdot \sigma_j ) \right\rangle
\end{eqnarray}
valid for $i \ne j$.
If we examine the behavior of the projected $Z(2)$ theory
in the vicinity of a second-order phase
transition of the underlying $O(N)$ model,
the correlation function equalities imply that the
critical index $\beta$ measured from the projected $Z(2)$ variable
correlation functions will be identical to the critical
indices of the underlying $O(N)$ model.
On the other hand, if the $O(N)$ model has a first-order transition,
a jump in the order parameter $\left\langle \sigma \right\rangle $
will cause a jump in $\left\langle \mu \right\rangle $ as well.
This is troubling, for the following reason.
Assume for the moment that the correlation functions of $\mu$
can be derived from some short-ranged effective $Z(2)$-invariant Hamiltonian,
whose parameters are smooth functions of the parameters of the
underlying Hamiltonian.
Standard universality arguments tell us that the critical
indices should be those of a $Z(2)$ model, rather than those
of an $O(N)$ model.
The success of projection means a clear failure for universality.
For example, $O(N)$ models are asymptotically free in
two dimensions, but the Ising model is not asymptotically free, and
has a second-order transition in two dimensions.

The conflict between projection and universality can be taken
over from spin systems to finite temperature gauge theories,
using the well-known arguments of Svetitsky and Yaffe.\cite{SvetYaffe}
The basic idea is that the critical behavior of a d-dimensional
gauge theory at finite temperature lies in the universality
class of (d-1)-dimensional spin systems which have a global symmetry
group equivalent to the center of the gauge group. For example,
the four-dimensional $SU(2)$ gauge theory has a finite temperature
deconfining transition which is in the universality class of the
three-dimensional Ising model. Projection to $U(1)$ must lead to
either a failure of universality or a failure of projection.
On the other hand, there is no difficulty whatsoever with
projection to $Z(2)$. In this sense, projection to the
center of the gauge group is more natural than projection to
other subgroups, such as the maximal Abelian subgroup.

There are three mechanisms by which this failure may be avoided.
The first two have been discussed
in the context of the renormalization group in the opus
by van Enter, Fernandez and Sokal.\cite{Sokal}
First, the parameters of the effective theory may not be well-behaved
functions of the underlying theory. Second, the effective
Hamiltonian may not exist.
Although the results of \cite{Sokal} are not directly applicable here,
it seems unlikely, based on their work, that the projection mapping
is discontinuous. 
It is possible that the effective Hamiltonian may not exist.
For an interesting example of what happens when a $Z(3)$ invariant system,
finite temperature $SU(3)$ lattice gauge theory, is reduced to
a $Z(2)$ spin system, the reader should consult the recent work
of Svetitsky and Weiss.\cite{SvetWeiss}
A third possibility is that the effective Hamiltonian may include
novel terms which remove it from its naive universality class.
The work by Yee on projection from $d=4$ 
$SU(2)$ lattice gauge theory to $U(1)$ is relevant here.\cite{Yee}
The $d=4$ $SU(2)$ model has no phase transition (at zero temperature),
but the standard $U(1)$ model does.
Using the demon method, Yee determined an approximate form for
the $U(1)$ effective action, showing that the projected theory
develops a magnetic monopole mass term which allows the effective
theory to avoid the phase transition of the standard $U(1)$ model.
Further work is needed here.

\section{\protect\vspace{1pt}Conclusions}

The success of Abelian projection appears to have its origin in very general
considerations. The key principle is local gauge invariance. Ultimately, it
is Elitzur's theorem\cite{Elitzur}
that ensures that observables constructed from the
projected field can always be rewritten in terms of gauge-invariant
observables of the underlying gauge fields. Abelian dominance is not
necessary. Note that at no point in the arguments given above has space-time
dimensionality been a consideration, a further indication that Abelian
projection does not depend on some particular set of important field
configurations.

\vspace{1pt}There is one possible weak point in the gauge-fixed case: the
standard gauge fixing algorithm corresponds formally to the limit $\lambda
\rightarrow \infty $, but the strong-coupling expansion in $\lambda $ has a
finite radius of convergence. Thus it is possible that these arguments fail
for large $\lambda $. Indeed, if we interpret the gauge-fixing field $\phi $
as a quenched adjoint scalar, it seems possible, based on studies of similar
models, that there is a phase transition along some critical line $\lambda
_{c}\left( \beta \right) $, a function of the gauge coupling $\beta $. The
nature of the phase transition will depend on the dimensionality of the
gauge-fixing functional; note that no phase transition will exist in the $d=1
$ case. If there is a phase transition in the gauge-variant sector, it may
be that the strong-coupling region and weak-coupling region are in fact
connected because the critical line has an end-point. Even if that is not
the case, it remains conceptually difficult to claim that confinement should
be understood differently for large $\lambda $ and small $\lambda $, because
by construction, the underlying ensemble of non-Abelian gauge fields does
not depend on $\lambda $. There is no grounds for saying one value
of $\lambda$ is more physical than another.
This is similar to continuum gauge fixing.
Although one may prefer Landau gauge to Feynman gauge for calculation,
one cannot say that Landau gauge is correct, but Feynman gauge is wrong.

Is there value in Abelian projection? At least some of its success appears
to have nothing to do with the dynamics of confinement, but rather follows
from general field-theoretic principles.
On the other hand,
Abelian projection may capture important aspects of the QCD\ vacuum,
although perhaps not in a unique way.
There is much we need to know:
1) Which features of projection follow from general principles?
We have shown here that the string tension is one such quantity.
It would be very interesting to extend the work presented here
on Wilson loops
to operators sensitive to
monopoles and other interesting topological objects.
2) Which features depend on the subgroup used for projection?
Although both center projection and projection to the maximal Abelian subgroup
can be used to obtain
the string tension, they differ in their approach
to confinement. Each has advantages over the other. For
example, center projection is consistent with universality,
but projection to the maximal Abelian subgroup is more easily
translated to the continuum.
3) Are there crucial tests for the various conceptual approaches
to projection and confinement which can be used to falsify them?
The ultimate
utility of Abelian projection can only be determined in conjunction with
theories which make testable quantitative predictions beyond the equality of
string tensions.

\section*{ACKNOWLEDGEMENTS}
I wish to thank Maarten Golterman, Sumit Das and especially Jeff
Greensite for useful discussions. I thank the U.S. Department of Energy
for financial support under grant number DE-FG02-91-ER40628.

\figure{ Labelling of sites and links for integration
over $\phi$ fields for $O(\lambda^3$ correction
to a Wilson loop.  \label{f1} }
\epsfysize=8in \epsfbox{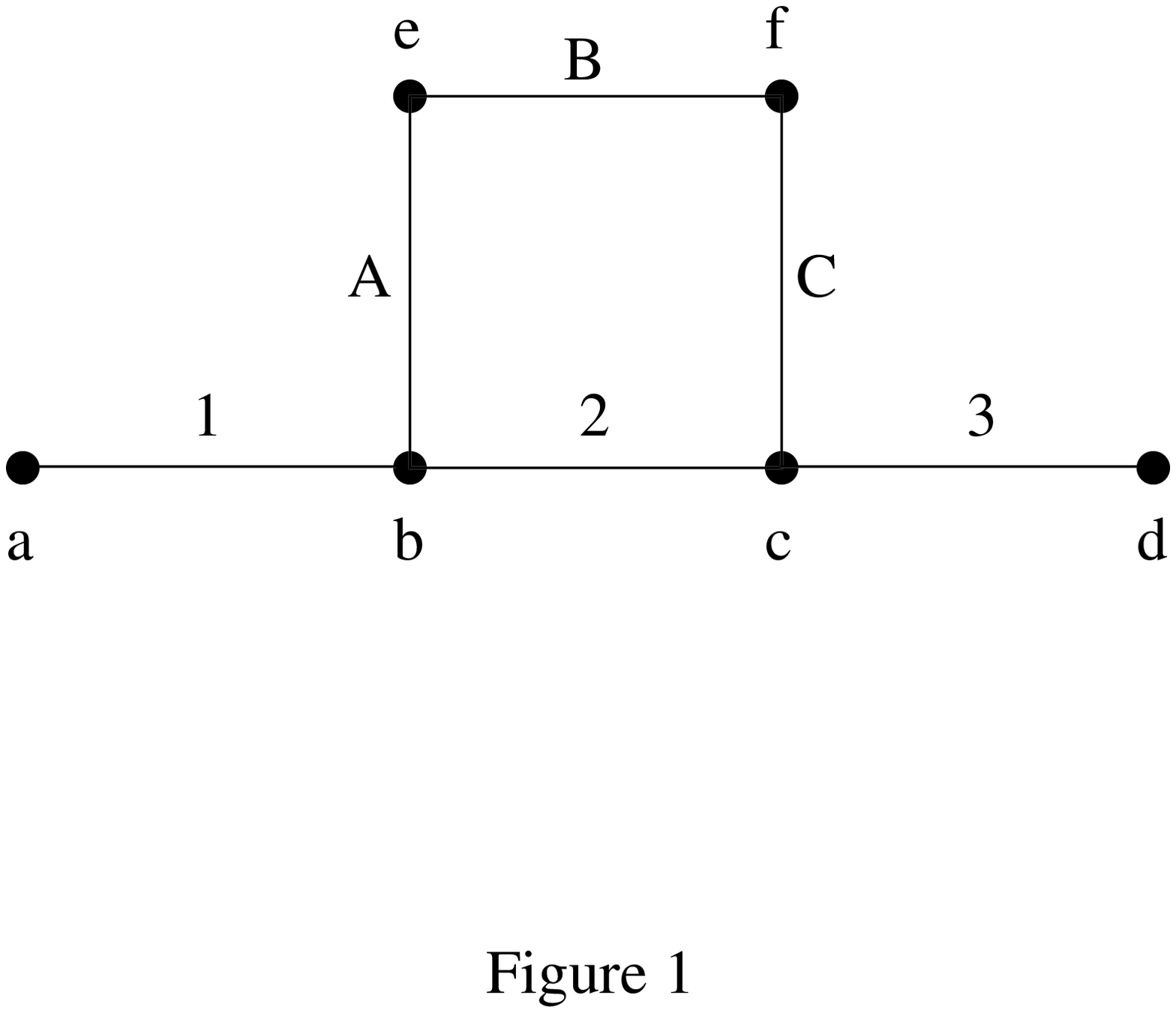}
\vspace{1.0in}
\pagebreak

\figure{ Approximate relation of $H$ Wilson loops to $G$ Wilson loops
with gauge fixing. The summation is over all possible handle
insertions.
\label{f2} }
\epsfysize=8in \epsfbox{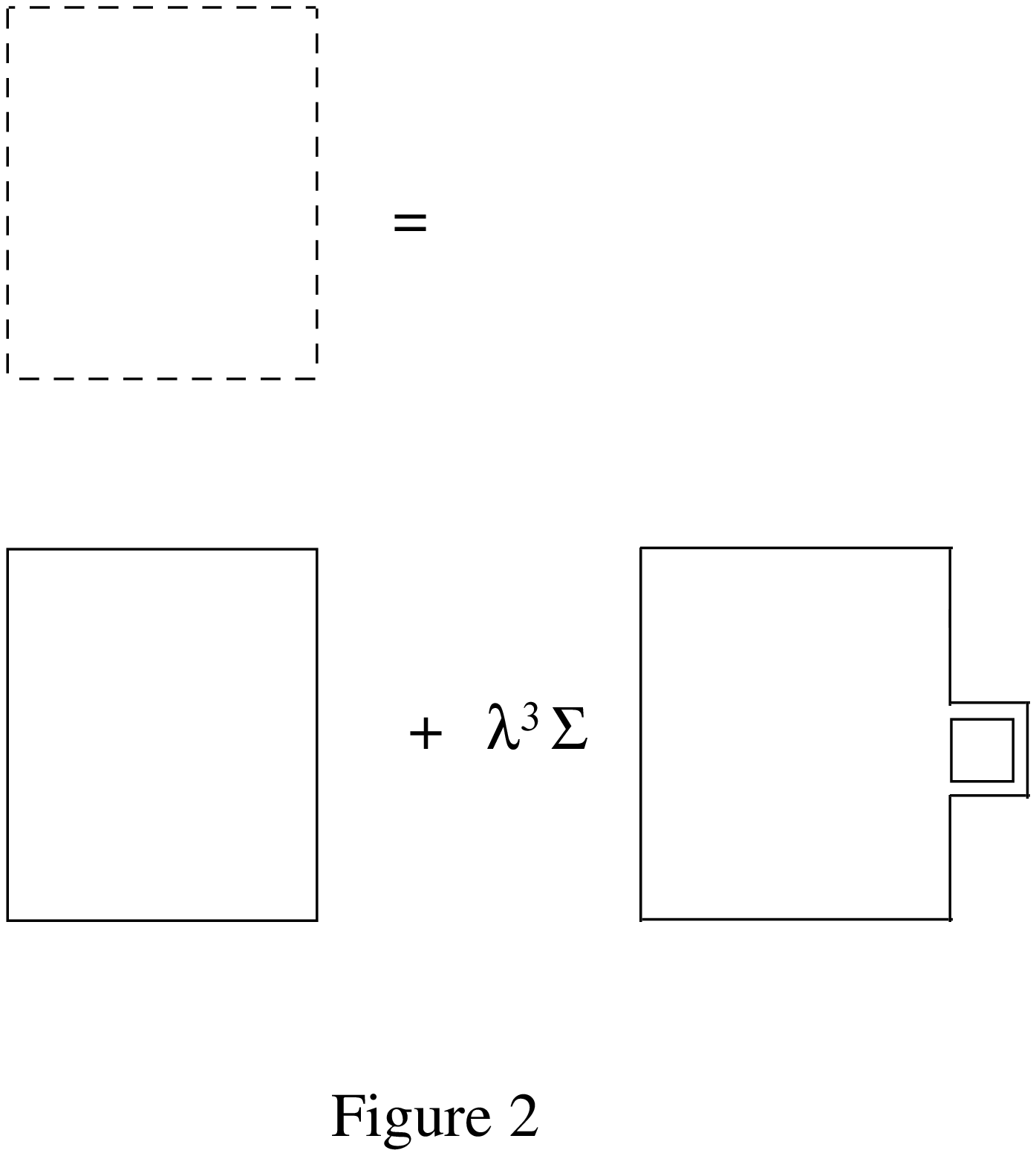}
\vspace{1.0in}
\pagebreak

\figure{ Approximate relation of $H$ Wilson loops to $G$ Wilson loops
with fat links. The summation is over all possible staple insertions.
\label{f3} }
\epsfysize=8in \epsfbox{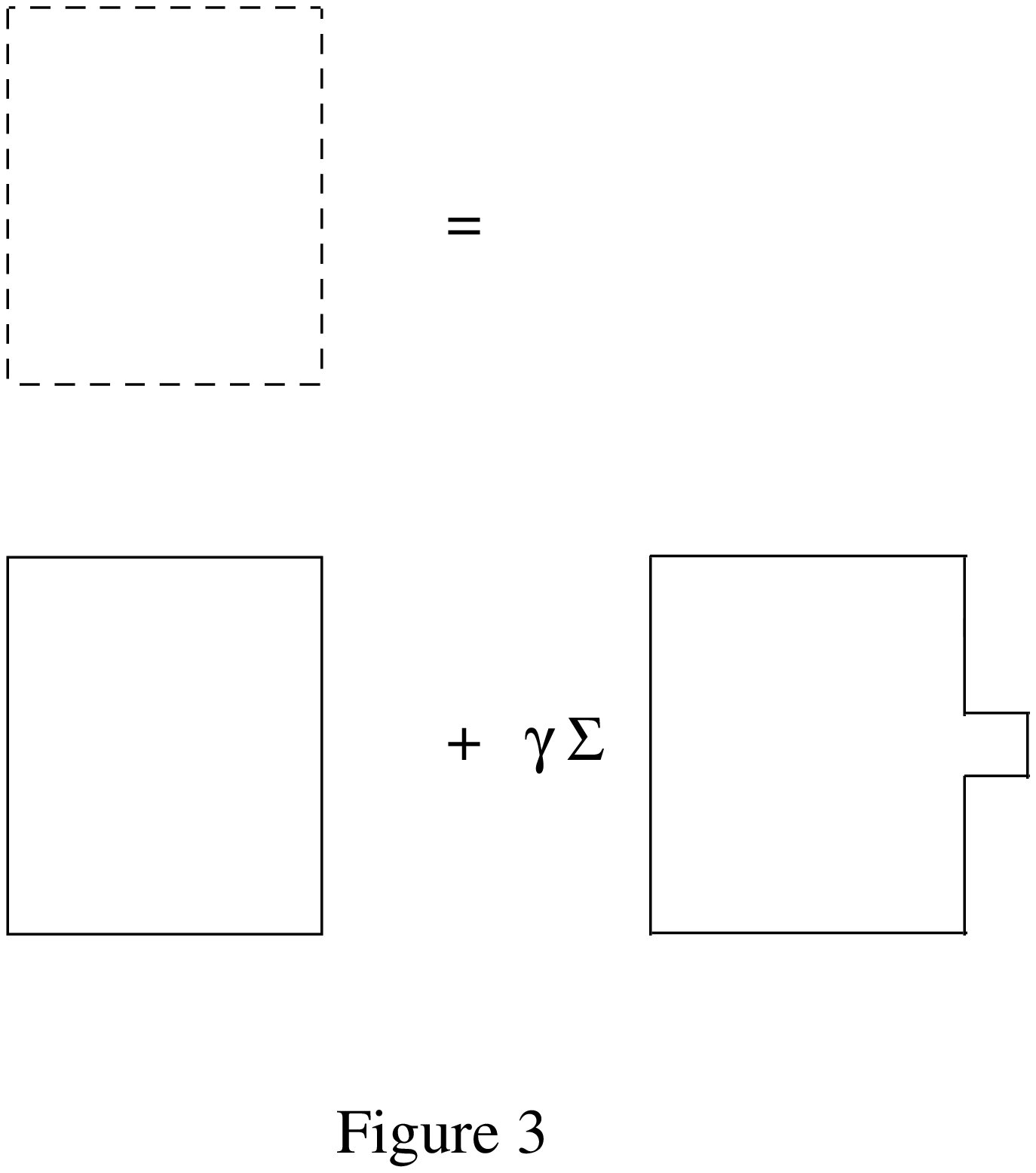}
\vspace{1.0in}
\pagebreak
\end{document}